\documentclass[12pt]{article}
\def\slash#1{\setbox0=\hbox{$#1$}#1\hskip-\wd0\hbox to\wd0{\hss\sl/\/\hss}}
\usepackage{epsf, cite, amsmath, amssymb}
\usepackage{epsfig}
\usepackage[all]{xy}
\usepackage{subfigure}
\usepackage{graphicx}

\makeatletter
\renewcommand\section{\@startsection {section}{1}{\z@}%
                                   {-3.5ex \@plus -1ex \@minus -.2ex}
                                    {2.3ex \@plus.2ex}%
                                   {\normalfont\large\bfseries}}
\renewcommand\subsection{\@startsection{subsection}{2}{\z@}%
                                     {-3.25ex\@plus -1ex \@minus -.2ex}%
                                     {1.5ex \@plus .2ex}%
                                     {\normalfont\bfseries}}
\makeatother

\let\non\nonumber

\newcommand{\bea}{\begin{eqnarray}}
\newcommand{\eea}{\end{eqnarray}}
\newcommand{\be}{\begin{equation}}
\newcommand{\ee}{\end{equation}}

\newcommand{\p}{\partial}

\newcommand{\td}{\text{D}}
\newcommand{\adsix}{\overline{\td6}}
\newcommand{\mR}{\mathbb{R}}

\newcommand{\C}[1]{$(\ref{#1})$}

\begin{document}

\begin{titlepage}

\begin{center}



\vskip 2 cm
{\Large \bf Generalized Gross--Neveu Models and Chiral Symmetry Breaking
from String Theory}\\
\vskip 1.25 cm { Anirban Basu$^{a}$\footnote{email: abasu@ias.edu}
and Anshuman Maharana $^{b}$\footnote{email: anshuman@physics.ucsb.edu}
}\\
{\vskip 0.75cm
$^a$ Institute for Advanced Study, Princeton, NJ 08540, USA\\
\vskip 0.2cm
$^b$ Department of Physics,
University of California Santa Barbara,\\ Santa Barbara, CA 93106, USA}

\end{center}

\vskip 2 cm

\begin{abstract}
\baselineskip=18pt

We consider intersecting D-brane models which have
two dimensional chiral fermions localized at the intersections. At weak coupling,
the interactions of these fermions are described by generalized Gross--Neveu models. At strong coupling,
these configurations are described by the dynamics of probe D-branes in a curved background
spacetime. We study patterns of dynamical chiral symmetry breaking in these models at weak and 
strong coupling, and also discuss relationships between these two descriptions.

\end{abstract}

\end{titlepage}

\pagestyle{plain}
\baselineskip=18pt

\section{Introduction}

Four dimensional Quantum Chromodynamics (QCD) describes the force
of strong interactions in nature. This theory is asymptotically
free and is strongly coupled at low energies, thus making it
difficult to analyze. In particular, it has proven difficult to
solve the theory even in the limit of large number of colors. In
the limit in which the quark masses in the QCD Lagrangian can be
neglected, this theory has a chiral flavor symmetry which is
broken in the presence of the mass terms. Understanding the
dynamics of chiral symmetry breaking in strongly coupled gauge
theories continues to be a challenge.

          An extremely successful phenomenological model to understand
chiral symmetry breaking in ${{\rm{QCD}}_4}$ was developed
originally by Nambu and Jona--Lasinio (NJL)~\cite{Nambu:1961tp}
where they constructed a field theory of chiral fermions with a
four--fermion interaction. In this model, the chiral flavor
symmetry gets dynamically broken to the diagonal subgroup and the
mesons are identified with the Nambu--Goldstone bosons of the
broken symmetry generators. However, this four--fermion
interaction is non--renormalizable and so the predictions depend
on the UV cutoff of the theory.

In two dimensions, an asymptotically free quantum field theory was constructed by Gross and
Neveu (GN)~\cite{Gross:1974jv}  which has a renormalizable four--fermion interaction. So the coupling
undergoes dimensional transmutation~\cite{Coleman:1973jx}, leaving the number of colors as the only free
parameter in the theory. This model can be exactly solved in the limit of large number of colors and
exhibits dynamical chiral symmetry breaking. Thus the GN model has proven to be an extremely interesting
toy model in understanding chiral symmetry breaking in QCD. It should be noted that the chiral symmetry
is broken only when the number of colors $N_c$ is strictly infinite, and is restored for any finite
value of $N_c$~\cite{Witten:1978qu,Affleck:1985wa}, however large. As described in~\cite{Witten:1978qu}, 
the $1/N_c$ expansion is
reliable in studying the spectrum of this model as well as related ones like the Thirring model.
So though we shall continue to use the term
``symmetry breaking'' as used in the literature, it is important to remember that the fermions are
massive, no continuous symmetries are broken, and the massless bosons are not
Nambu--Goldstone bosons.

 It is believed that string theory techniques will play
an important role in understanding strong coupling issues like
confinement and chiral symmetry breaking in QCD. A confining pure
Yang--Mills $U (N_c)$ gauge theory in four dimensions was
constructed in~\cite{Witten:1998zw} by compactifying one of the
world--volume dimensions of $N_c$ D4-branes on a circle of radius
$\text{R}$ with supersymmetry breaking boundary conditions. The
fermions and the scalars of the world volume theory of the
D4-branes get masses at tree level and one--loop level
respectively, and so the infra--red theory is pure Yang--Mills
which exhibits confinement. In order to model more realistic
theories involving chiral fermions, the authors
of~\cite{Sakai:2004cn,Sakai:2005yt} considered a D--brane
configuration involving ``flavor'' D8-branes alongwith the
``color'' D4-branes, where the flavor branes intersect the color
branes along three spatial dimensions, such that there are no
directions transverse to both the flavor and the color branes. The
flavor branes are separated by a distance $L$ along the D4
world--volume. One gets chiral fermions in this model which are
given by the 4--8 strings stretching between the flavor and color
branes. These chiral fermions are localized at the intersections
of the color and flavor branes. Though classically this
configuration preserves the chiral flavor symmetry, this model
exhibits dynamical chiral symmetry breaking (which is a global
symmetry from the point of view of the color brane theory). This
was demonstrated at large values of a classically dimensionless
coupling constant $\lambda/L$ (where $\lambda$ is essentially the
't Hooft coupling of five dimensional Yang--Mills theory )
in~\cite{Sakai:2004cn}, by considering the D8-branes as
probes~\cite{Karch:2002sh}\footnote{Also
see~\cite{Son:2003et,Kruczenski:2003uq,Babington:2003vm} for
related discussions.} in the near--horizon geometry of the $N_c$
D4-branes. Chiral symmetry breaking manifests itself as a wormhole
solution of the D8-brane action connecting the
D8-$\overline{\text{D}8}$ pair.

Now in these models when one takes the transverse separation
$L$ to be of the same order of magnitude as the circle
radius $R$, the scales of chiral symmetry breaking and confinement
are comparable and it becomes harder to analyze the dynamics of
chiral symmetry breaking without taking into account the effects
of confinement, and vice versa. Thus to look at the dynamics when
chiral symmetry is broken but the theory is not confining, one can
consider the limit when $R \rightarrow \infty$. This has been done
in~\cite{Antonyan:2006vw}\footnote{The issue of separation of
scales was discussed in a different context in~\cite{Bak:2004nt}.}
and the configuration has been analyzed in the limits $\lambda /L
\rightarrow 0$ and $\lambda /L \rightarrow \infty$, and it has
been found that the chiral symmetry is broken in both the regimes.
The $\lambda /L \rightarrow 0$ regime is described by a non--local
version of the NJL model (see~\cite{Volkov:2005kw} for a review of
phenomenological applications of the non--local NJL model), while
the opposite regime is described by a wormhole configuration
similar to that mentioned above. It has also been conjectured that
these solutions lie in the same universality class as
${\rm{QCD}_4}$ which is obtained by taking $R$ to be finite.

     A similar analysis to study chiral symmetry breaking in two dimensions
has been done in~\cite{Antonyan:2006qy}. The D-brane configuration
is given by the color D4-branes which intersect the flavor
D6-branes along one spatial dimension such that there are no
directions transverse to both the color and the flavor branes.
Dynamical breaking of chiral symmetry occurs both in the $\lambda
/L \rightarrow 0$ and $\lambda /L \rightarrow \infty$ limits. The
$\lambda /L \rightarrow 0$ regime is described by the GN model,
while the opposite regime is described by a wormhole
configuration. This setup is expected to be in the same
universality  class as $\rm{QCD}_2$, which is obtained by wrapping
three of the world--volume directions of the D4-branes on $T^3$.

      In this paper, we generalize the construction of~\cite{Antonyan:2006qy} 
by considering multiple stacks of ``flavor'' D6 (and 
$\overline{\td6}$)-branes which intersect the ``color'' D4-branes
along one spatial direction. The flavor branes are placed at various
points in $\mR^3$ which parametrizes the world--volume directions of the 
D4-branes that are transverse to the intersection. Our motivation is 
two--fold: to understand the pattern of chiral symmetry breaking in multi-brane
constructions, and to study the relationship between the weak and
strong coupling descriptions. In the limit $\lambda /L_i
\rightarrow 0$ (where $L_i$ refers to the distances between the
various D6 and $\overline{\text{D}6}$ pairs in $\mR^3$) the
D-brane configurations reduce to generalized GN models\footnote{GN
models with more than one coupling have been considered, for
example, in~\cite{Klimenko:1985ss}.}, where we exhibit various
patterns of chiral symmetry breaking. In the opposite regime, we
describe these configurations in terms of probe D6-branes in the
near--horizon geometry of the color branes. We find a close
relationship between the two descriptions.

            The plan of the paper is as follows. We begin by describing
the D-brane setup in the next section, followed by the analysis
of various setups where all the flavor branes are collinear in the transverse
$\mR^3$. These configurations have some interesting features, and
we discuss the relationship between chiral symmetry breaking in the strong 
and weak coupling limits. We next generalize this construction to flavor branes
spanning an $\mR^2$ in $\mR^3$, and we finally consider general
flavor brane configurations in $\mR^3$.

        The pattern of chiral symmetry breaking has distinct
features. In all our examples we find that at weak coupling, not
all the possible condensates are non--vanishing. The strong
coupling analogue of condensates is the presence of  wormholes
connecting the brane and anti-brane pairs. Just like at weak coupling, the 
wormhole configuration is also determined by the energetics, and leads to patterns 
of chiral symmetry breaking similar to weak coupling.

       Finally we describe patterns of restoration of the chiral
symmetries as the temperature of the system is raised. At a
sufficiently high temperature, all the symmetries are restored.

\section{The D-brane setup }

                               The GN model can be realized in
string theory \cite{Antonyan:2006qy} by considering
$N_{c}$ ``color'' D4-branes extending along the (01234) directions
and two stacks of ``flavor'' D6 and $\overline{\text{D}6}$-branes which
extend in the directions (0156789). To discuss generalizations of
the GN model, we shall consider multiple stacks of D6 and
$\overline{\text{D}6}$-branes given by

\begin{equation}
\nonumber
\begin{array}{cccccccccccccccccccccccccccccccccccccccc}
  &   &0& \ \  &1& \ \ &2& \ \ &3& \ \ &4& \ \  &5& \
\ &6& \ \ &7&  \ \ &8&  \ \ &9& \\
\\
 \text{D}4& :  &\textsf{x}& \  \ &\textsf{x}& \  \ &\textsf{x}& \
     \ &\textsf{x}&\  \ &\textsf{x}&  \\
 \\
 \text{D}6\text{s} ,\ {\overline{\text{D}6}\text{s}}&  : &\textsf{x}& \ \
&\textsf{x}& \ \ & \ & \ \ & \ & \ \ & \ \ & \ \ &\textsf{x}& \
\ &\textsf{x}& \ \ &\textsf{x}&  \ \ &\textsf{x}&  \ \ &\textsf{x}& \\
\\
\end{array}
\label{intersection}
\end{equation}

All the flavor branes extend
in the (0156789) directions, and are located at different
points in $\mR^3$ spanned by the
directions (234).
We shall take every stack of flavor branes to be composed of $N_{f}$
branes, our results can be easily generalized to the case where
the stacks consist of different numbers of branes.

             It is straightforward to deduce the massless spectrum
corresponding to these D-brane configurations. Apart from the
obvious $U(N_c)$ gauge theory on the D4-brane world--volume, and
the $U(N_f)_i$ gauge theory on the world--volume of the $i-$th
stack of $N_f$ D6 (or $\overline{\td6}$)-branes, there are extra
normalizable massless modes coming from the 4-6 strings that
stretch between the D6 ($\overline{\td6}$)-branes and the
D4-branes. These massless modes are spacetime fermions coming from
the Ramond sector, the lowest modes in the Neveu--Schwarz sector
are massive. These 4-6 strings are localized in the directions
(01) and give rise to chiral fermions in two dimensions. Thus
every stack of D6-branes gives rise to left--moving fermions $q_L$
which transform in the $(N_f, N_c)$ of $U(N_f)_L \times U(N_c)$.
Similarly, every stack of $N_f$ $\overline{\td6}$-branes produces
right--moving fermions $q_R$ which transform in the $(N_f, N_c)$
of $U(N_f)_R \times U(N_c)$. Thus the spectrum of a configuration
consisting of m stacks of D6 and n stacks of
$\overline{\td6}$-branes has m left--movers $q_L$ and n
right--movers $q_R$. There is a $U(N_{f})$ symmetry associated
with each stack of flavor branes. The fermions are charged under
the flavor group of the stack they are localized on, and transform
trivially under the flavor group of other stacks.

    We will be always working in the limit where $\alpha' \rightarrow
0$, $g_s \rightarrow 0$, $N_c \rightarrow \infty$, with $g_s N_c$
and $N_f$ fixed\footnote{Thus, we will not be considering
processes involving the annihilation of the branes and the
anti-branes.}. In this limit, the gauge coupling of the flavor
branes vanish compared to that of the color branes, so the
only relevant coupling is the 't Hooft coupling 
\begin{equation}
 \lambda = \frac{g_5^2 N_c}{4\pi^2},
\end{equation}
where $g_5^2 = 4\pi^2 g_s \alpha'^{1/2}$ is the dimensionful 
coupling of the five dimensional Yang--Mills theory. As in~\cite{Antonyan:2006qy},
we shall discuss the physics of the
system in two tractable regimes given by the classically
dimensionless couplings $\lambda /L_i \rightarrow 0$ and
$\lambda /L_i \rightarrow \infty$, where $L_i$ are the distances
between the D6-$\overline{\td6}$ pairs.

  In the former limit, the system is well described by an
interacting theory of the chiral fermions with the D4-brane gauge
field. These fermions are derivatively coupled to the D4-brane
scalars and so these couplings vanish in the $\alpha' \rightarrow
0$ limit. As we shall demonstrate below, one can integrate out the
gauge field along the lines of~\cite{Antonyan:2006qy} to obtain
generalized GN models.
The symmetries associated with the flavor stacks $U(N_f)_L$ for the
left--moving fermions, and $U(N_f)_R$ for the right--moving ones appear
as global symmetries.

                     In the
later limit $\lambda /L_i \rightarrow \infty$, the 't Hooft
coupling $\lambda$ is large and/or the flavor branes are close
to each other. Then
the five--dimensional gauge theory effects are large, and we
cannot use the above description. However we now have an alternate
weakly coupled description~\cite{Itzhaki:1998dd} which involves
analyzing the D6-brane dynamics in the near--horizon geometry of
the color D4-branes, which we will also discuss below.

\section{Three collinear flavor branes}

      We begin with simplest non--trivial examples of chiral symmetry
breaking which involve three stacks of flavor branes placed along
a straight line in the $\mR^3$ spanned by the directions (234).
These will illustrate the general features of symmetry breaking
and vacuum structure in our models. We shall rely heavily on these
configurations while discussing later the general flavor brane
setups in $\mR^3$. With three collinear branes, there are two
physically distinct orderings\footnote{Other configurations are
related to the ones we discuss by charge conjugation.}, given by
$\overline{\td6}-\td6-\overline{\td6}$ and
$\td6-\overline{\td6}-\overline{\td6}$ which we discuss below. In
what follows, it will be useful to introduce indices on the stacks
of flavor branes. We label the $\td6$-brane stack by $1$ and the
$\adsix$-brane stacks by labels $\bar{1}$ and $\bar{2}$.

\subsection{ $\overline{\td6}-\td6-\overline{\td6}$ }

    Consider a stack of $\td6$-branes and two stacks of $\overline{\td6}$-branes
placed along the $4$ direction in the order
$\overline{\td6}-\td6-\overline{\td6}$ as shown in figure \C{fig1}. 
The stack of D6-branes is located at
the origin while the $\overline{\td6}$-brane stacks have coordinates $L_{1}$
and $-L_{2}$ in the $4$ direction.

\begin{figure}[ht]
\begin{center}
\[
\mbox{\begin{picture}(240,125)(0,0)
\includegraphics[scale=.5]{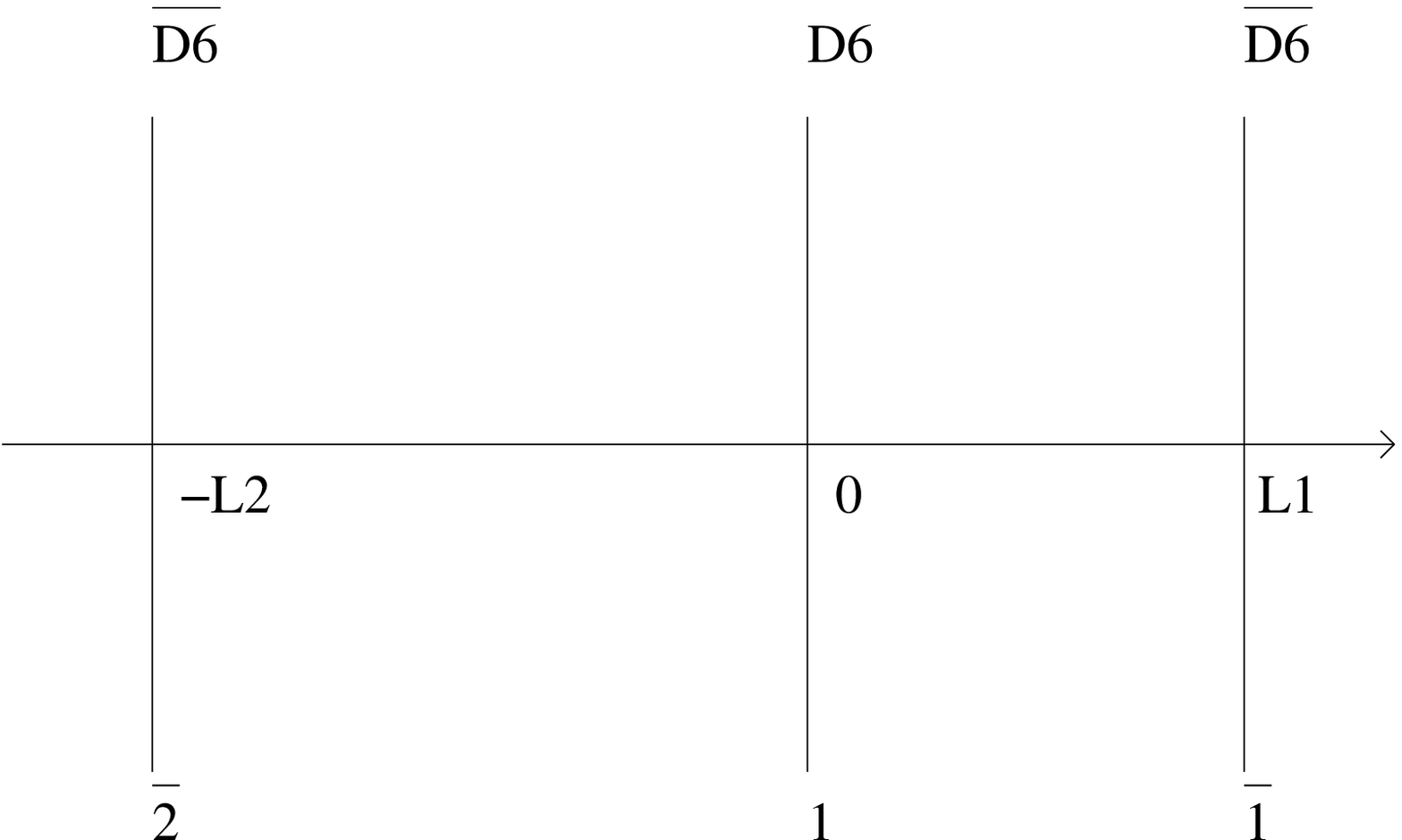}
\end{picture}}
\]
\caption{ \label{fig1} Collinear $\overline{\td6}-\td6-\overline{\td6}$. }
\end{center}
\end{figure}

     In the regime where the separation  between the stack of
D6-branes and the stacks of $\overline{\td6}$-branes is much larger
compared to the five dimensional 't Hooft coupling $\lambda$, the effective action
is given by
\begin{eqnarray}
 S &=& \int d^5 x \Big[ -\frac{1}{4g_5^2} F_{MN}^2 + \delta^3
(\vec{x}) q_{1}^\dagger(i\p_+ + A_+)  q_{1} + \delta^3 (\vec{x}
-L_1 \hat{x}_4) q_{\bar{1}}^\dagger (i\p_- + A_-) q_{\bar{1}} \non
\\&& + \delta^3 (\vec{x} + L_2 \hat{x}_4) q_{\bar{2}}^\dagger
(i\p_- + A_-) q_{\bar{2}} \Big] \, \label{fst},
\end{eqnarray}
where $q_{1}$ is the left--handed fermion localized at stack 1,
and $q_{\bar{1}}$ and $q_{\bar{2}}$ are the right--handed fermions localized
at stacks $\bar{1}$ and $\bar{2}$ respectively. Also $F_{MN}$, $M,N = 0,1,2,3,4$,
is the field strength for the five dimensional gauge field $A_M$, and $A_\pm = A_0 \pm A_1$
are its light--cone components.

   To leading order in $\lambda / L_{i}$, the dynamics of the fermions
can be studied by integrating out the five dimensional gauge field
in the single gluon exchange approximation~\cite{Antonyan:2006qy}.
This can be easily done in Feynman gauge, and one obtains an action
with a four-fermion interaction

\begin{eqnarray} \label{nonlocGN} S &=& \int d^2 x \Big[ q_{1}^\dagger i\p_+
q_{1}  + {q}_{\bar{1}}^\dagger i\p_- {q}_{\bar{1}} + {q}_{\bar{2}}^\dagger i\p_- {q}_{\bar{2}} \Big] \non \\
&&+\frac{g_5^2}{4\pi^2} \int d^2 x d^2 y \Big[ G (x-y, L_1)
\Big( q_{1}^\dagger (x) \cdot  q_{\bar{1}} (y) \Big) \Big( q_{\bar{1}}^\dagger (y)
\cdot q_{1} (x) \Big)  \non
\\ &&+ G (x-y, L_2)
\Big( q_{1}^\dagger (x) \cdot  q_{\bar{2}} (y) \Big) \Big( q_{\bar{2}}^\dagger (y)
\cdot q_{1} (x) \Big) \Big],
\end{eqnarray}
where $G(x,L)$ is the Euclidean propagator
\begin{equation}
 G(x, L) = \frac{1}{(x^2 +
L^2)^{3/2}}.
\end{equation}
In the four--fermion interactions, the dot represents color index
contractions, while the flavor indices are contracted in the
obvious way. This action is a generalization of the non--local GN
model. Now we can further consider the local limit of this model
where the fields do not fluctuate over distances of order $L_i$,
and we consider the theory at distance scales much larger then $L_i$~\footnote{As
in~\cite{Antonyan:2006qy}, one can continue to work with the
non--local model, however in the limit we work in, it gives the
same result.}. In this limit the propagator essentially behaves as
a delta function smeared over distances $L_i$, and using
\begin{equation}
 \int d^2 x G(x,L) = \frac{2\pi}{L}
\end{equation}
in (\ref{nonlocGN}), we get a generalized local GN model with
action

\begin{eqnarray} \label{def1}
  S &=& \int
d^2 x \Big[ q_{1}^\dagger i\p_+ q_{1}
+ {q}_{\bar{1}}^\dagger i\p_- {q}_{\bar{1}} + {q}_{\bar{2}}^\dagger i\p_- {q}_{\bar{2}} \Big] \non \\
&&+ \frac{1}{ N_c}\int d^2 x \Big[  \frac{2\pi \lambda}{L_1}
(q_{1}^\dagger \cdot  q_{\bar{1}}) (q_{\bar{1}}^\dagger \cdot
q_{1}) + \frac{2\pi \lambda}{L_2} (q_{1}^\dagger \cdot
q_{\bar{2}}) (q_{\bar{2}}^\dagger \cdot q_{1})  \Big] \ .
\end{eqnarray}
Note that the couplings of the four-fermion interactions are given
by the ratios of the five dimensional 't Hooft coupling and the
separations between the branes and the anti-branes.

   This model can be exactly solved in the large $N_c$ limit. We briefly
outline the steps involved in analyzing the vacuum structure. We
begin by introducing auxiliary bosonic fields and writing the
action as
\begin{eqnarray}
 S &=& \int d^2 x \Big[  q_{1}^\dagger
i\p_+ q_{1} + q_{\bar{1}}^\dagger i\p_- q_{\bar{1}} + q_{\bar{2}}^\dagger i\p_- q_{\bar{2}}\Big] \non \\
&&+\int d^2 x \Big[ -\frac{1}{N_c}\frac{L_1 }{2\pi \lambda} \vert
\phi_1 \vert^2 -\frac{1}{N_c}\frac{L_2}{2\pi \lambda} \vert
\phi_2 \vert^2 + (\overline\phi_1 q_{{1}}^\dagger \cdot
q_{\bar{1}} + \overline\phi_2 q_{{1}}^\dagger \cdot  q_{\bar{2}} +
{\rm {h.c.}}) \Big].
\end{eqnarray}
In the large $N_{c}$ limit, the effective potential can be
explicitly evaluated by integrating out the fermions\footnote{We quote
results for $N_{f}=1$, these
generalize in a straightforward manner for arbitrary $N_{f}$.}. Using
dimensional regularization, we get 
\begin{equation}
\frac{V_{\rm{eff}}}{N_c} =  \frac{L_{1} }{ 2\pi \lambda} \vert
\phi_1 \vert^2 + \frac{ L_{2} }{2\pi \lambda} \vert \phi_2
\vert^2 + \frac{\vert \phi_1 \vert^2 + \vert \phi_2
\vert^2}{4\pi}\Big[{\rm ln} \Big( \frac{\vert \phi_1 \vert^2 +
\vert \phi_2 \vert^2}{\mu^2} \Big) -1 \Big], \label{pot}
\end{equation}
where $\mu$ is the renormalization scale.

             The effective potential (\ref{pot}) has three extrema
\begin{equation} \label{extri}
\begin{array}{ccccccccccccccccccccccccccccccccccccccccccccccccc}
& \ \ \      & \vert \phi_1 \vert &  \ \ \ \ \ \ \ \  \ \ \ \ & \vert \phi_{2} \vert & \ \ \ \ \ \ \ \
&V_{\rm{eff}} \\
\\
&\text{A.}  \ \ \   &\mu e^{-L_{1}/ \lambda}&       &0&
&-\frac{\mu^{2}N_c}{4\pi}e^{-2L_{1}/\lambda} \\
\\
&\text{B.}  \ \ \        &0&       &\mu e^{-L_{2}/ \lambda}&
&-\frac{\mu^{2}N_c}{4\pi}e^{-2L_{2}/\lambda} \\
\\
&\text{C.} \ \ \ &0& \ \ \ &0& \ \ \ &0
\end{array}
\end{equation}

           While C is the global maximum, the global minimum is determined by the relative
magnitudes of $L_{1}$ and $L_{2}$. Without loss of generality, we
take $L_{1} < L_{2}$ for our discussion\footnote{We shall treat
the case $L_{1}=L_{2}$ separately.}, and thus the extremum A 
corresponds to the vacuum. The field $\phi_{1}$ which corresponds
to the fermion bilinear $q_{1}^\dagger \cdot q_{\bar{1}}$, acquires
a non--vanishing vacuum expectation value, while $\phi_{2}$ does
not condense. As a result the classical chiral symmetry $U
(N_{f})_1 \times U (N_{f})_{\bar{1}} \times U (N_{f})_{\bar{2}}$
is dynamically broken to $U (N_{f})_{\rm{diag}(1,\bar{1})} \times
U (N_{f})_{\bar{2}}$. The fermions $q_{1}$ and $q_{\bar{1}}$ acquire
mass much smaller than the energy scale
$\mu$~\cite{Gross:1974jv}, while $q_{\bar{2}}$ continues to be massless.
Note that the extremum B is tachyonic (along the $\phi_{1}$
direction) with mass
\begin{equation}
 \label{tacheqn} m^2 = \frac{\p^2 V_{\rm{eff}}}{\p \phi_1 \p
{\bar\phi}_1} = \frac{2\pi}{\lambda} (L_1 -L_2) <0.
\end{equation}
Many of these statements shall have interesting counterparts in
the strong coupling regime, which we discuss next.

 When $\lambda /L_i$ is large, the configuration admits a weakly coupled dual
description~\cite{Antonyan:2006qy} which we briefly review. 
Consider a stack of D6-branes
and a stack of $\overline{\td6}$-branes located at $\pm
\frac{L}{2}$ along the $4$ direction respectively as shown in
figure \C{fig2}.
\begin{figure}[ht]
\begin{center}
\[
\mbox{\begin{picture}(260,125)(0,0)
\includegraphics[scale=.5]{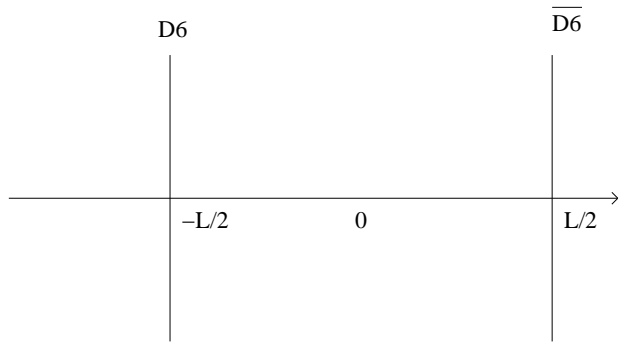}
\end{picture}}
\]
\caption{ \label{fig2} D6-$\overline{\td6}$ pair. }
\end{center}
\end{figure}
To study the behavior of the system at strong coupling, consider a
probe D6-brane in the near--horizon geometry of $N_c$ D4-branes,
which is given by the metric and the dilaton
\begin{eqnarray}
  ds^{2} &=& \bigg( \frac{ U}{R} \bigg)^{3/2}\big[ \eta_{\mu \nu}dx^{\mu}
  dx^{\nu} - (dx^{4})^{2} \big] - \bigg( \frac{U}{R}
  \bigg)^{-3/2}(dU^{2} + U^{2}d \Omega_{4}^{2}), \non \\
  e^{\Phi} &=& g_{s} \bigg( \frac{U}{R} \bigg)^{3/4}, \non
\end{eqnarray}
where
\begin{equation}
R^3 =\pi \lambda ,
\end{equation}
$U$ is the radial coordinate, and $\Omega_{4}$ labels the angular directions 
in (56789). The analysis
in~\cite{Antonyan:2006qy} revealed a solution where the D6-brane
extends in the (01) directions, wraps the four sphere labeled by
$\Omega_{4}$ and has a wormhole like profile $U(x_{4})$ which
asymptotes
\begin{figure}[ht]
\begin{center}
\[
\mbox{\begin{picture}(165,125)(-15,-15)
\includegraphics[scale=.5]{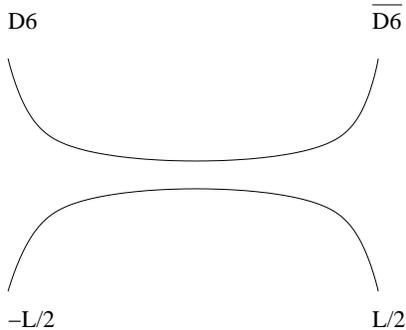}
\end{picture}}
\]
\caption{ \label{fig3} The wormhole. }
\end{center}
\end{figure}
to the undeformed D6 and $\overline{\td6}$-branes at infinity as
shown in figure \C{fig3}. The wormhole connecting the branes manifestly
exhibits chiral symmetry breaking. In fact, one can show that this
configuration has less energy than the undeformed
D6--$\overline{\td6}$ pair with

\begin{equation}
 \label{minen}
  \delta E (L) \approx - \frac{\lambda^2}{L^4}.
\end{equation}

                   Our configuration admits two wormhole solutions, because
the D6-branes in stack 1 can connect onto either of the two stacks of
$\adsix$-branes at $\bar{1}$ and $\bar{2}$ (see figures \C{fig4} and \C{fig5}).

\begin{figure}[ht]
\begin{center}
\[
\mbox{\begin{picture}(140,125)(-15,-15)
\includegraphics[scale=.5]{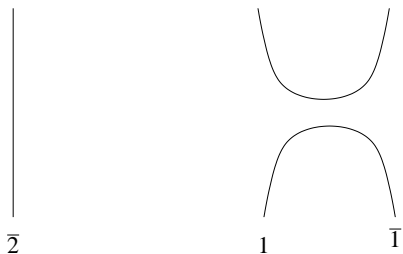}
\end{picture}}
\]
\caption{\label{fig4} Vacuum configuration. }
\end{center}
\end{figure}

\begin{figure}[ht]
\label{twoways}
\begin{center}
\[
\mbox{\begin{picture}(140,125)(-15,-15)
\includegraphics[scale=.5]{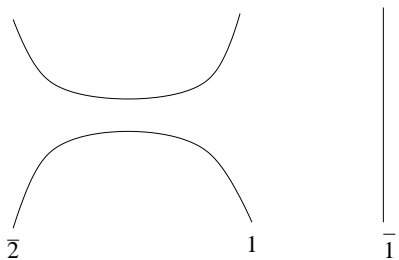}
\end{picture}}
\]
\caption{\label{fig5} Metastable configuration. }
\end{center}
\end{figure}

      From (\ref{minen}) we see that energetics requires that the D6-branes 
in stack 1 connect
to the closer $\overline{\td6}$-brane stack at $\bar{1}$, while 
$\bar{2}$ remains disconnected, as depicted in
figure \C{fig4}. Thus as in the weak coupling regime, the classical
chiral symmetry $U (N_{f})_1 \times U (N_{f})_{\bar{1}} \times U
(N_{f})_{\bar{2}}$ is broken to $U (N_{f})_{\rm{diag} (1,\bar{1})}
\times U (N_{f})_{\bar{2}}$.

   Note that the strong coupling counterpart of vanishing $\phi_{2}$
condensate is the absence of a wormhole connecting stacks 1 and $\bar{2}$. This
illustrates an interesting point about general D-brane configurations at strong
coupling. At strong coupling, all ``condensates'' except those corresponding to
the pairs of branes and anti-branes which are connected by
wormholes vanish. In all examples we will consider later in the
paper, we find that the weak coupling vacuum structure has
behavior which seems to be reminiscent of this property.

                                The energetically disfavored
configuration in figure \C{fig5} has the same symmetries as the
extremum B in (\ref{extri}) and it is natural to think of it as
the strong coupling continuation of B. We note that the corresponding
wormhole configuration is metastable, while B in \C{extri} has a tachyon
given by \C{tacheqn} .

  Finally, we consider the particular case when $L_1 = L_2 \equiv L$.
In the weak coupling regime, from \C{nonlocGN} we see that the
non--local GN model (and consequently the local GN model) has an
enhanced $U (N_f)_L \times U (2 N_f)_R$ chiral symmetry under which

\begin{displaymath}{q_{\bar{1}} \choose q_{\bar{2}}}
\end{displaymath}
transforms in the $2N_f$ dimensional representation of $U (2 N_f)_R$. Thus
\begin{displaymath}{\phi_1 \choose \phi_2}
\end{displaymath}
also transforms in the same way. In this case, the vacuum
configuration of the GN model is given by 

\be  \vert \phi_1
\vert^2 + \vert \phi_2 \vert^2 = \mu^2 e^{-2L/\lambda},
\ee
which leads to the vacuum energy

\be V_{\rm{eff}} = -\frac{\mu^2 N_c}{4\pi} e^{-2L/\lambda} .\ee

Thus the $U (N_f)_L \times U (2 N_f)_R$ chiral symmetry is dynamically broken to
$U (N_f)_{\rm{diag} (L,R)} \times  U (N_f)_R$. At strong coupling, this 
symmetry breaking manifests itself in the fact that the wormhole
can connect the D6-brane with either of the two
$\overline{\td6}$-branes as they both have the same energy.

\subsection{ $\td6-\overline{\td6}-\overline{\td6}$}

          The other configuration involves placing the branes and
anti-branes in the order $\td6-\overline{\td6}-\overline{\td6}$ as
shown in figure \C{fig6}.
\begin{figure}[ht]
\begin{center}
\[
\mbox{\begin{picture}(240,125)(0,0)
\includegraphics[scale=.5]{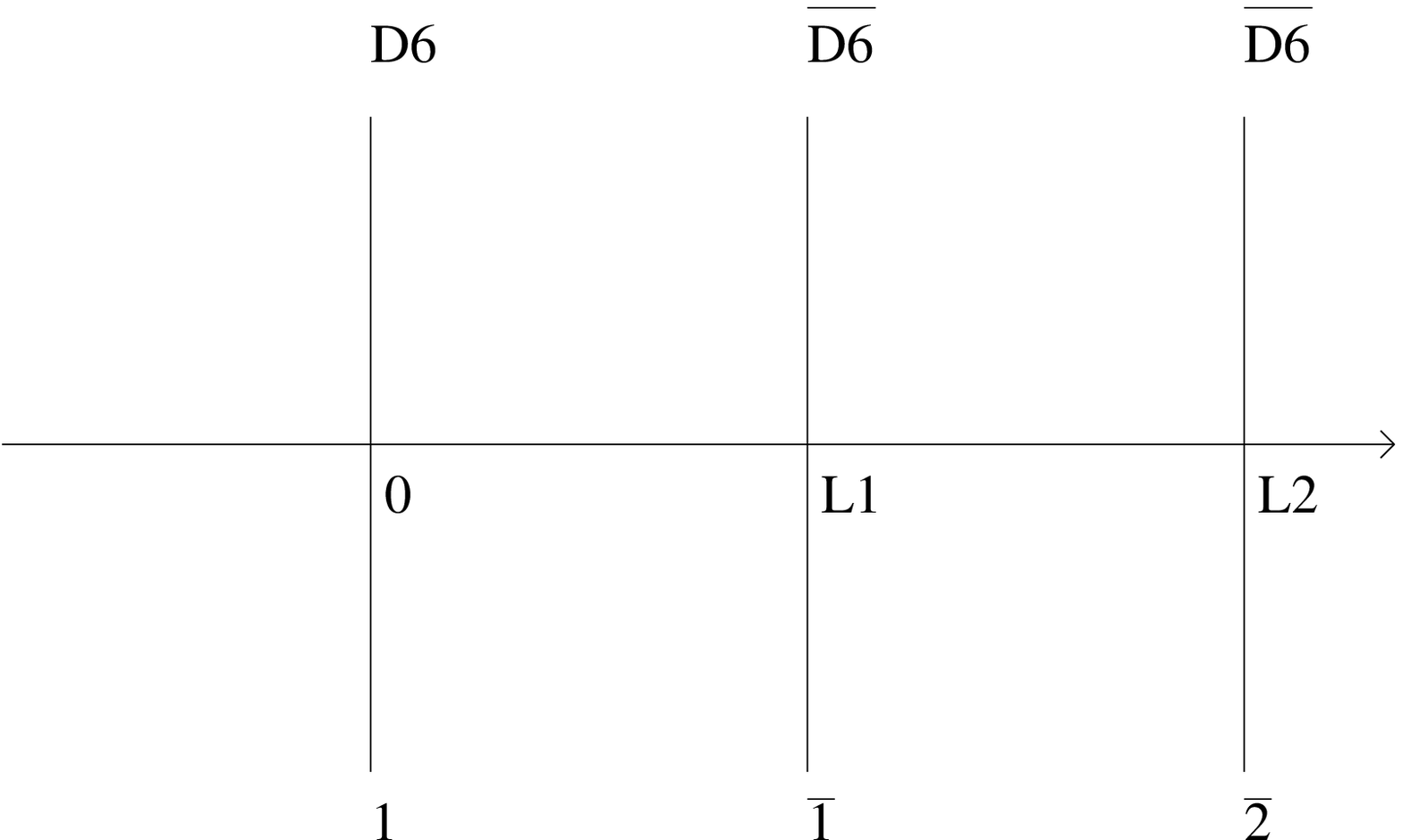}
\end{picture}}
\]
\caption{\label{fig6} Collinear $\td6-\overline{\td6}-\overline{\td6}$.
}
\end{center}
\end{figure}

We place the stack of D6-branes at the origin and the two stacks of
$\overline{\td6}$-branes at $L_{1}$ and $L_{2}$, where $L_2 > L_1$.
The analysis can be easily carried out making making use of
the results of the previous section. At weak coupling, the
couplings between the left and right-handed fermions depend only
of the distances between the brane and anti-brane pairs. Hence the
vacuum energy density is identical to the previous case and is
given by (\ref{pot}). From (\ref{extri}) we see that $\phi_{1}$ has
a non-vanishing vacuum expectation value, and so the classical
$U (N_{f})_1 \times U (N_{f})_{\bar{1}} \times U
(N_{f})_{\bar{2}}$ chiral symmetry is dynamically broken to $U
(N_{f})_{\rm{diag}(1,\bar{1})} \times U (N_{f})_{\bar{2}}$. At
strong coupling, the D6-brane is connected to the anti
$\overline{\td6}$-brane closer to it through a wormhole as
shown in figure \C{fig7}.

\begin{figure}[ht]
\begin{center}
\[
\mbox{\begin{picture}(140,125)(-15,-15)
\includegraphics[scale=.5]{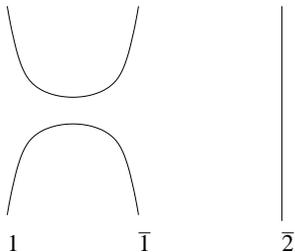}
\end{picture}}
\]
\caption{\label{fig7} $U (N_{f})_1 \times U (N_{f})_{\bar{1}} \times U
(N_{f})_{\bar{2}} \rightarrow U
(N_{f})_{\rm{diag}(1,\bar{1})} \times U (N_{f})_{\bar{2}}$. }
\end{center}
\end{figure}

\section{More patterns of chiral symmetry breaking}

Having analyzed chiral symmetry breaking in three collinear stacks of flavor branes,
we now consider more general patterns involving D-brane configurations which
have equal numbers of D6 and $\overline{\td6}$-branes. First, we shall consider four
collinear stacks of flavor branes. Then we shall consider two simple D-brane
configurations spanning an $\mR^2$ in $\mR^3$. Apart from being useful models
to study patterns of chiral symmetry breaking, these constructions will be helpful
in analyzing the general pattern of symmetry breaking involving arbitrary configurations
of flavor branes in $\mR^3$. As before we find that
the patterns of chiral symmetry breaking are similar at weak and strong
coupling.

\subsection{Four collinear flavor branes}

       In this section we shall consider configurations with two
stacks of D6-branes, and two stacks of $\adsix$-branes placed along the 4 direction. We shall
label the D6-brane stacks by indices $1,2$, and denote their coordinates by $L_{1},L_{2}$
respectively. Similarly, we denote the $\adsix$-brane stacks by
$\bar{1}, \bar{2}$, and denote their coordinates by
$L_{\bar{1}},L_{\bar{2}}$ respectively. We also use $L_{i \bar{j}}$ to denote
the distance between the $i$-th $\td6$-brane and the $\bar{j}$-th $\adsix$-brane.

     We first consider the weak coupling limit and obtain the effective potential.
As in the three stack case, to leading order in
$\lambda / L_{i \bar{j}}$, the dynamics of the fermions at length
scales much greater than the D-brane separations can be described by a local GN model obtained
by integrating out the five dimensional gauge field in the single gluon exchange approximation.
Once again, the effective field theory is a generalized GN model with two
left--moving and two right--moving fermions, and is given by

\begin{eqnarray}
  S = \int
d^2 x  \Big[ \sum_i q_{i}^\dagger i\p_+ q_{i} +
\sum_{\bar{j}} {q}_{\bar{j}}^\dagger i\p_- {q}_{\bar{j}}  +  \frac{1}{N_c}
\sum_{i,\bar{j}}
\frac{2\pi \lambda}{L_{i \bar{j}} } (q_{i}^\dagger \cdot
q_{\bar{j}}) (q_{\bar{j}}^\dagger \cdot q_{i})  \Big] \ .
\end{eqnarray}

After introducing auxiliary fields $\phi_{i \bar{j}}$
corresponding to fermion bilinears $q^{\dagger}_{i} \cdot q_{\bar{j}}$
and integrating out the fermions, one obtains the effective potential
in the large $N_c$ limit

\begin{eqnarray}
\frac{V_{\rm{eff}}}{N_c} =   \sum_{i, \bar{j}} \frac{L_{i
\bar{j}}}{2\pi \lambda} \vert \phi_{i \bar{j}} \vert^2   + \frac{
\theta + \sqrt{\delta}}{8\pi}\Big[{\rm ln} \Big( \frac{\theta +
\sqrt{\delta}}{2\mu^2} \Big) -1 \Big]   + \frac{  \theta -
\sqrt{\delta}}{8\pi} \Big[ {\rm ln} \Big( \frac{ \theta -
\sqrt{\delta}}{2\mu^2} \Big) - 1 \Big], \label{fved}
\end{eqnarray}
where
\begin{equation}
     \theta = \sum_{i, \bar{j}} | \phi_{i \bar{j}}|^{2} \, ,\ \ \ \delta =
\Big( \sum_{i, \bar{j}} | \phi_{i \bar{j}}|^{2} \Big)^2
-4 | \phi_{1\bar{1}} \phi_{2\bar{2}}
     - \phi_{1\bar{2}} \phi_{2\bar{1}}|^{2}.
\end{equation}

 The properties of the vacuum configuration obtained by minimizing (\ref{fved})
are closely related to the ordering of the stacks (i.e., the relative
magnitudes of $L_{i\bar{j}}$). Upto charge conjugation, there are
three distinct orderings  $\td6-\td6-\adsix-\adsix$,
$\td6-\adsix-\adsix-\td6$ and $\td6-\adsix-\td6-\adsix$. We
discuss each case separately, also illustrating the relationship to
wormhole solutions at strong coupling. \\

\subsubsection{$\td6-\td6-\overline{\td6}-\overline{\td6}$}

\begin{figure}[ht]
\begin{center}
\[
\mbox{\begin{picture}(240,125)(0,0)
\includegraphics[scale=.5]{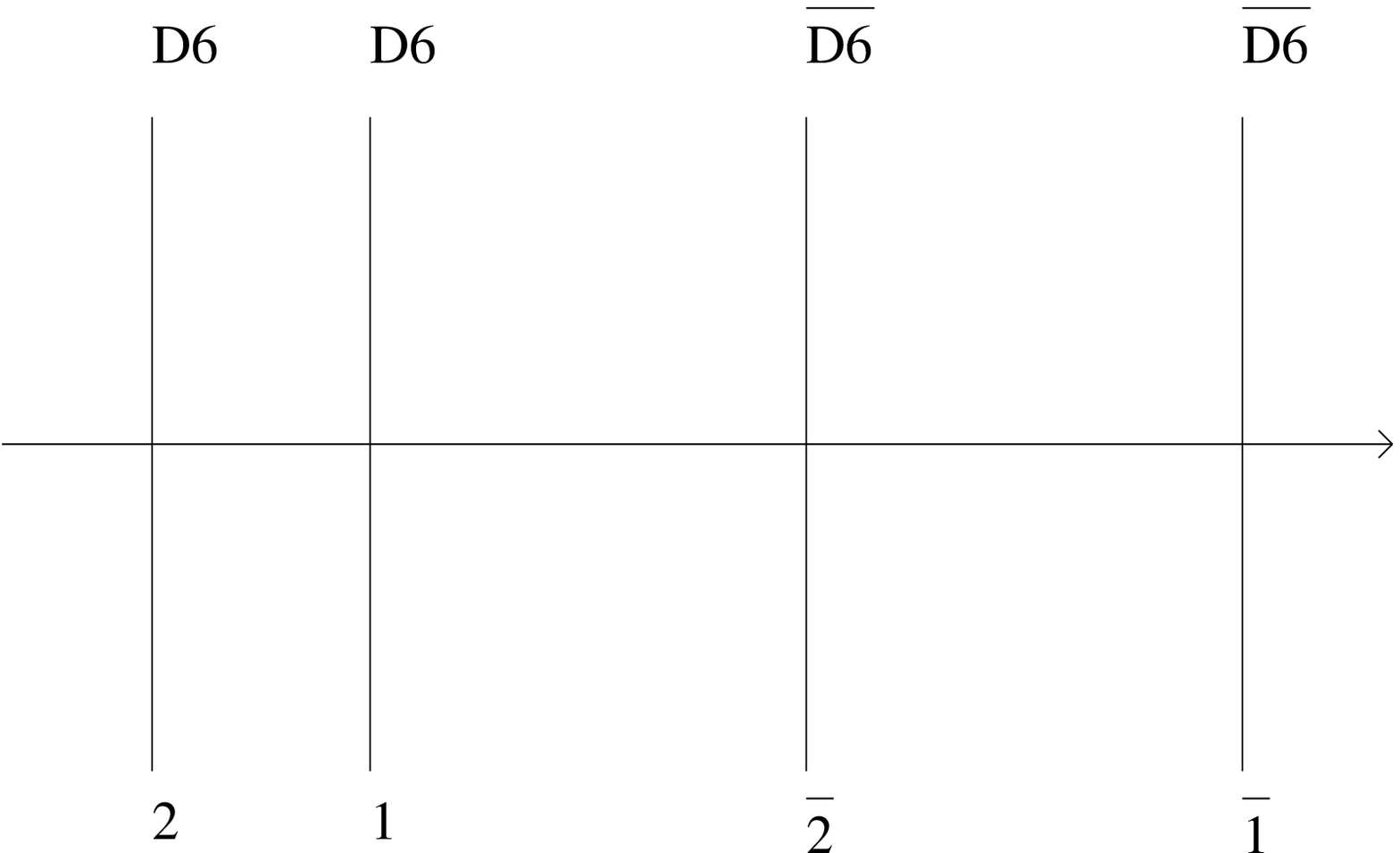}
\end{picture}}
\]
\caption{\label{fig8} Collinear $\td6-\td6-\overline{\td6}-\overline{\td6}$.
}
\end{center}
\end{figure}
\noindent         For this ordering given by figure \C{fig8}, the minimum of the effective 
potential (\ref{fved}) is given by
\begin{equation}
     \vert \phi_{1\bar{2}} \vert
= \mu e^{ -L_{1 \bar{2}}/ \lambda} \,, \ \ \  \vert \phi_{2\bar{1}} \vert
= \mu e^{ -L_{2 \bar{1}}/ \lambda} ,\ \
      \ \ \phi_{1\bar{1}} =0, \ \  \phi_{2\bar{2}}  = 0 \ .
\end{equation}
Thus the classical $U (N_{f})_1 \times U (N_{f})_2
\times U (N_{f})_{\bar{1}} \times
U (N_{f})_{\bar{2}}$ chiral symmetry is broken to
$U (N_{f})_{\rm{diag} (1,\bar{2})} \times
U (N_{f})_{\rm{diag} (2,\bar{1})}$. At strong coupling the wormhole
configuration with the lowest energy is shown in figure \C{fig9},
which exhibits the same symmetry breaking
pattern\footnote{The difference between the radii of the two
wormholes at the point of closest approach is much larger than
$\sqrt{\alpha'}$ for sufficiently large values of $\lambda/L_{i\bar{j}}$,
thus the gravity construction can be trusted.}.

\begin{figure}[ht]
\begin{center}
\[
\mbox{\begin{picture}(140,100)(0,0)
\includegraphics[scale=.5]{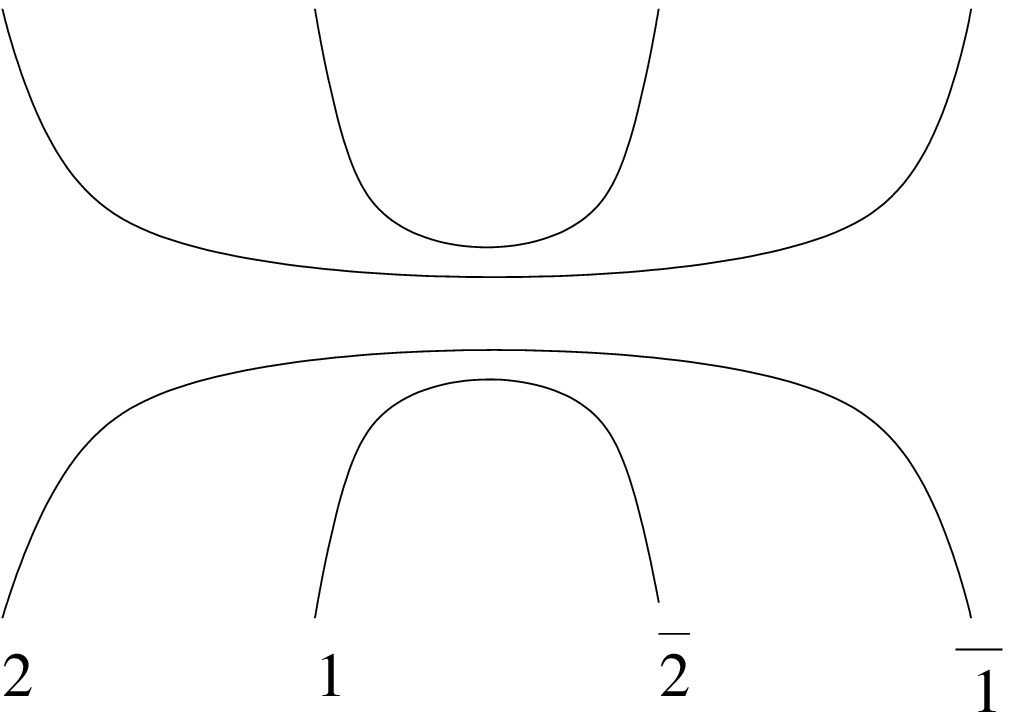}
\end{picture}}
\]
\caption{\label{fig9} $U (N_{f})_1 \times U (N_{f})_2
\times U (N_{f})_{\bar{1}} \times
U (N_{f})_{\bar{2}} \rightarrow U (N_{f})_{\rm{diag} (1,\bar{2})} \times
U (N_{f})_{\rm{diag} (2,\bar{1})}$. }
\end{center}
\end{figure}

\subsubsection{$\td6-\overline{\td6}-\overline{\td6}-\td6$}

\begin{figure}[ht]
\begin{center}
\[
\mbox{\begin{picture}(240,125)(0,0)
\includegraphics[scale=.5]{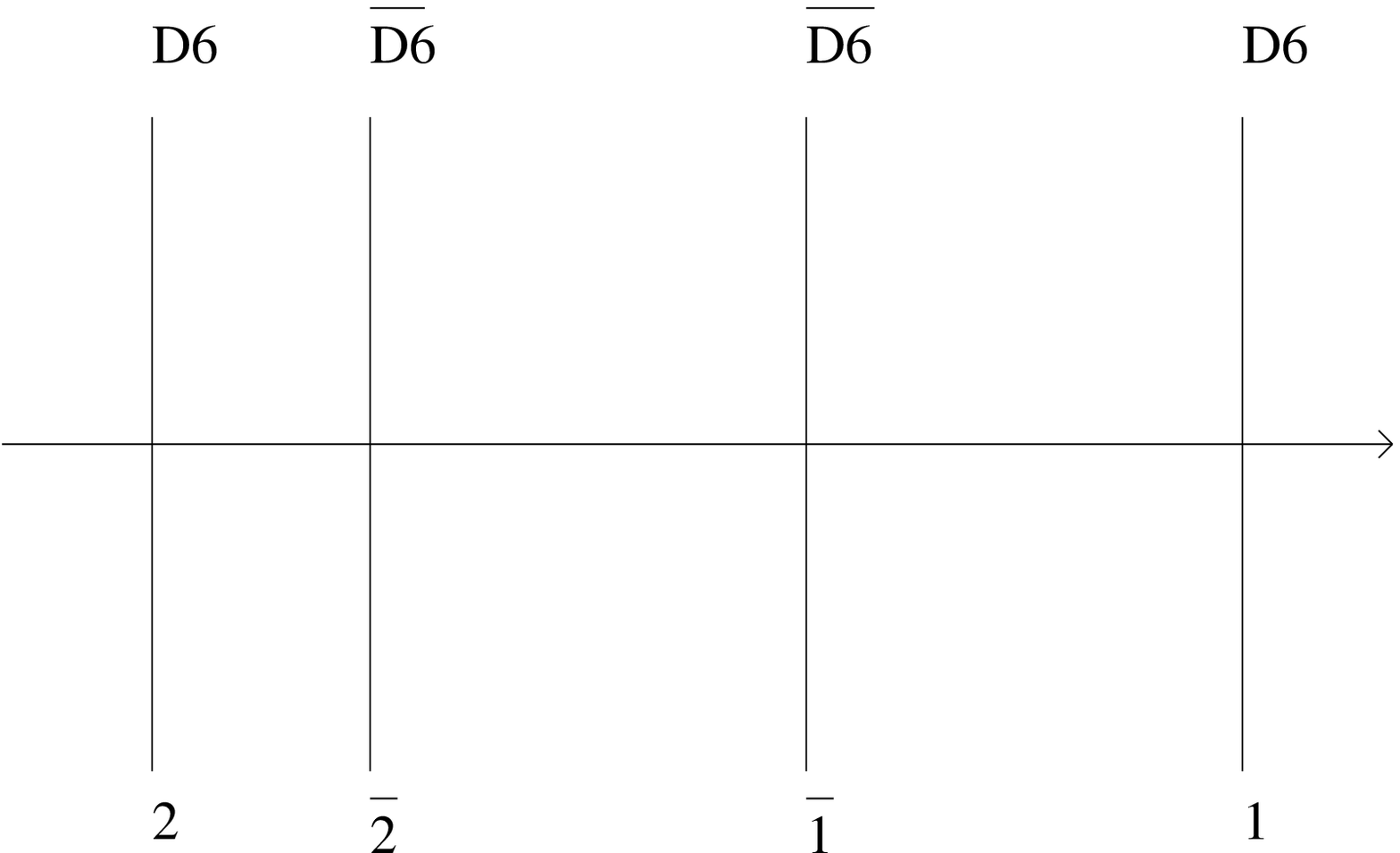}
\end{picture}}
\]
\caption{\label{fig10} Collinear $\td6-\overline{\td6}-\overline{\td6}-\td6$.
}
\end{center}
\end{figure}

 In this case given by figure \C{fig10}, the minimum of the effective potential (\ref{fved}) is given by
\begin{equation}
     \vert  \phi_{1\bar{1}} \vert
= \mu e^{ -L_{1 \bar{1}}/ \lambda} \, ,\ \ \  \vert \phi_{2\bar{2}} \vert
= \mu e^{ -L_{2 \bar{2}}/ \lambda} ,\ \
      \ \ \phi_{1\bar{2}} =0, \ \  \phi_{2\bar{1}}  = 0 \ ,
\end{equation}
and the chiral symmetry is dynamically broken to
$U (N_{f})_{\rm{diag} (1,\bar{1})} \times
U (N_{f})_{\rm{diag} (2,\bar{2})}$.
Again, the energetically favorable wormhole configuration is shown in figure \C{fig11},
which exhibits the same symmetry breaking pattern.

\begin{figure}[ht]
\begin{center}
\[
\mbox{\begin{picture}(225,125)(-15,-15)
\includegraphics[scale=.5]{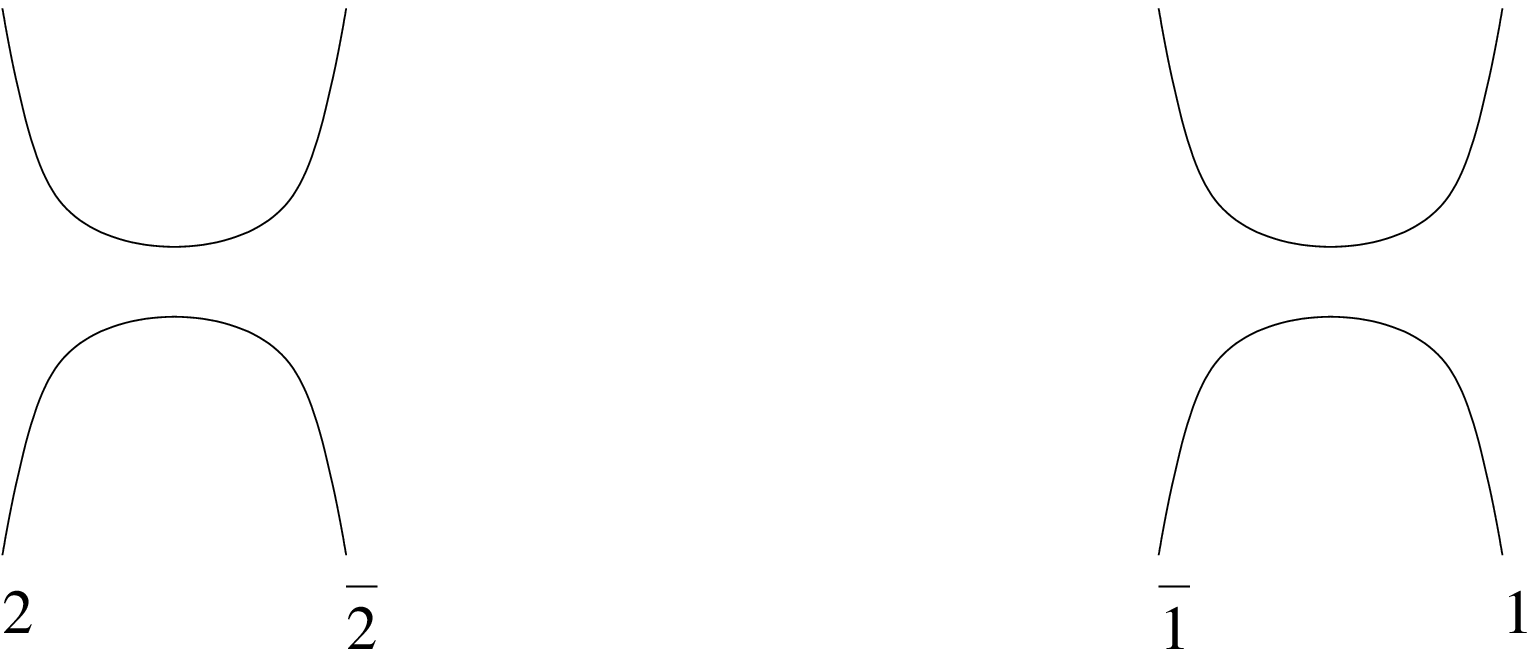}
\end{picture}}
\]
\caption{\label{fig11} $ U (N_{f})_1 \times U (N_{f})_2
\times U (N_{f})_{\bar{1}} \times
U (N_{f})_{\bar{2}} \rightarrow U (N_{f})_{\rm{diag} (1,\bar{1})} \times
U (N_{f})_{\rm{diag} (2,\bar{2})}$.
}
\end{center}
\end{figure}

\subsubsection{$\td6-\overline{\td6}-\td6-\overline{\td6}$}

\begin{figure}[ht]
\begin{center}
\[
\mbox{\begin{picture}(240,125)(-15,-15)
\includegraphics[scale=.5]{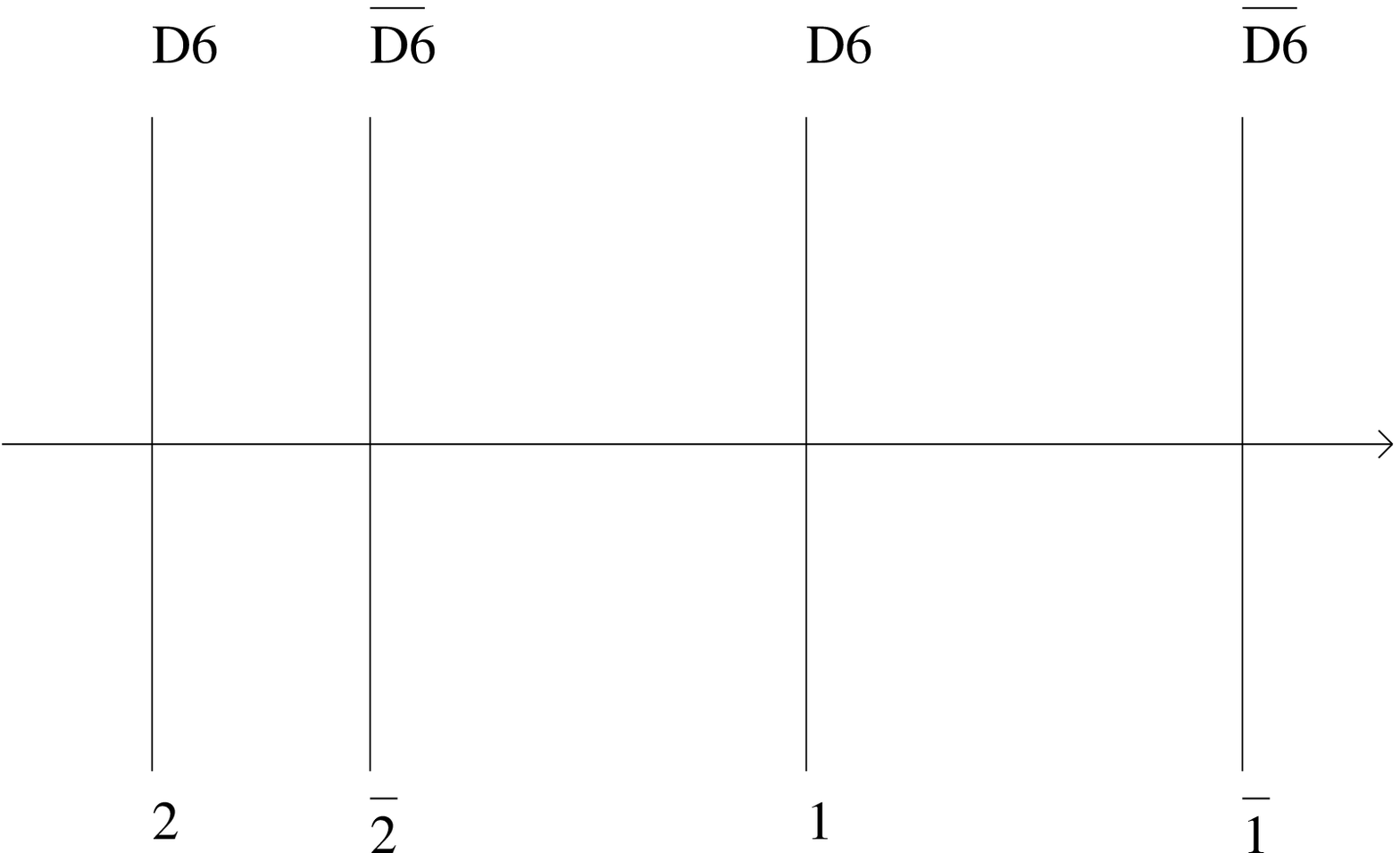}
\end{picture}}
\]
\caption{\label{fig12} Collinear $\td6-\overline{\td6}-\td6-\overline{\td6}$.
}
\end{center}
\end{figure}

   This case as shown in figure \C{fig12} exhibits more interesting structure of chiral symmetry breaking. 
The nature
of the vacuum structure explicitly depends on the separations of the stacks. At weak coupling,
simple energetics
using the effective potential (\ref{fved}) shows that for
\begin{equation}
 e^{-2L_{1 \bar{1}} / \lambda } +  e^{-2L_{2 \bar{2}} / \lambda}
> \ e^{-2L_{1 \bar{2}} /\lambda } +  e^{-2L_{2 \bar{1}} / \lambda},
\label{cri}
\end{equation}
the vacuum has condensates
\begin{eqnarray}
 \text{X} \ : \ \vert \phi_{1\bar{1}}\vert
 = \mu e^{ -L_{1 \bar{1}}/ \lambda} \,, \ \ \  \vert \phi_{2\bar{2}} \vert
= \mu e^{ -L_{2 \bar{2}}/ \lambda} ,\
      \ \ \phi_{1\bar{2}} =0, \ \  \phi_{2\bar{1}}  = 0. \
\label{phasea}
\end{eqnarray}
On the other hand, when the inequality in (\ref{cri}) is reversed
the condensates are
\begin{equation}
 \text{Y} \ : \ \vert \phi_{1\bar{2}} \vert
= \mu e^{ -L_{1 \bar{2}}/ \lambda} \, ,\ \ \  \vert \phi_{2\bar{1}} \vert
= \mu e^{ -L_{2 \bar{1}}/ \lambda}, \ \
      \ \ \phi_{1\bar{1}}=0, \ \  \phi_{2\bar{2}}  = 0 .\
\label{phaseb}
\end{equation}
Note that X and Y represent two distinct phases, with different
symmetries of the vacuum.

Similarly in the strong coupling regime, there are two patterns of
chiral symmetry breaking determined by the energetics of the wormhole configurations.
For

\begin{equation}
\label{ineq2}
 \frac{1}{L^{4}_{1 \bar{1}}} + \frac{1}{L_{2
\bar{2}}^4}
> \frac{1}{L_{1 \bar{2}}^4} + \frac{1}{L_{2 \bar{1}}^4},
\end{equation}
the energetically favored configuration is shown in figure \C{fig13}. On the other hand,
the configuration in figure \C{fig14} is energetically favored when the inequality in
(\ref{ineq2}) is reversed. The configurations in figures \C{fig13} and \C{fig14}
preserve the same symmetries as the phases X and Y respectively. They
should be thought of as the strong coupling continuations of these phases.

\begin{figure}[ht]
\begin{center}
\[
\mbox{\begin{picture}(225,125)(-15,-15)
\includegraphics[scale=.5]{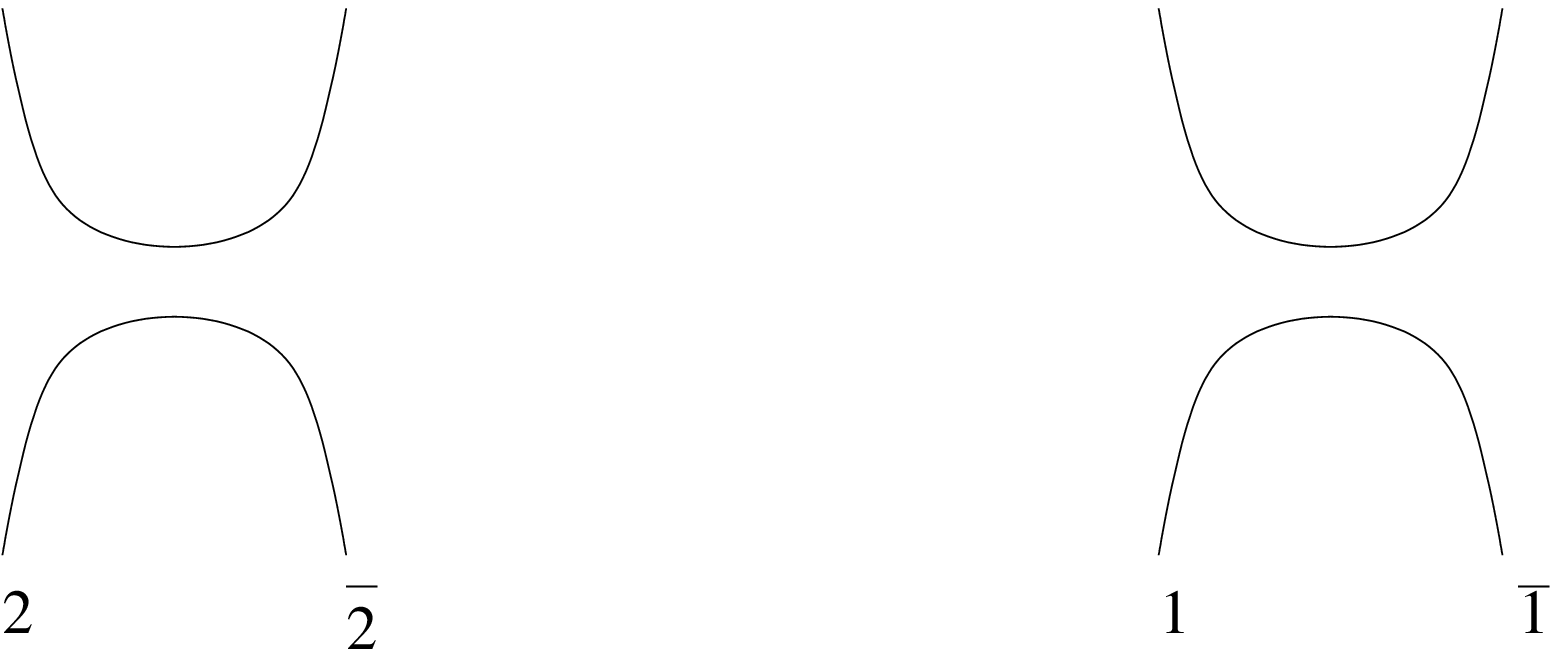}
\end{picture}}
\]
\caption{\label{fig13} $ U (N_{f})_1 \times U (N_{f})_2
\times U (N_{f})_{\bar{1}} \times
U (N_{f})_{\bar{2}} \rightarrow U (N_{f})_{\rm{diag} (1,\bar{1})} \times
U (N_{f})_{\rm{diag} (2,\bar{2})}$.
}
\end{center}
\end{figure}

\begin{figure}[ht]
\begin{center}
\[
\mbox{\begin{picture}(140,125)(-15,-15)
\includegraphics[scale=.5]{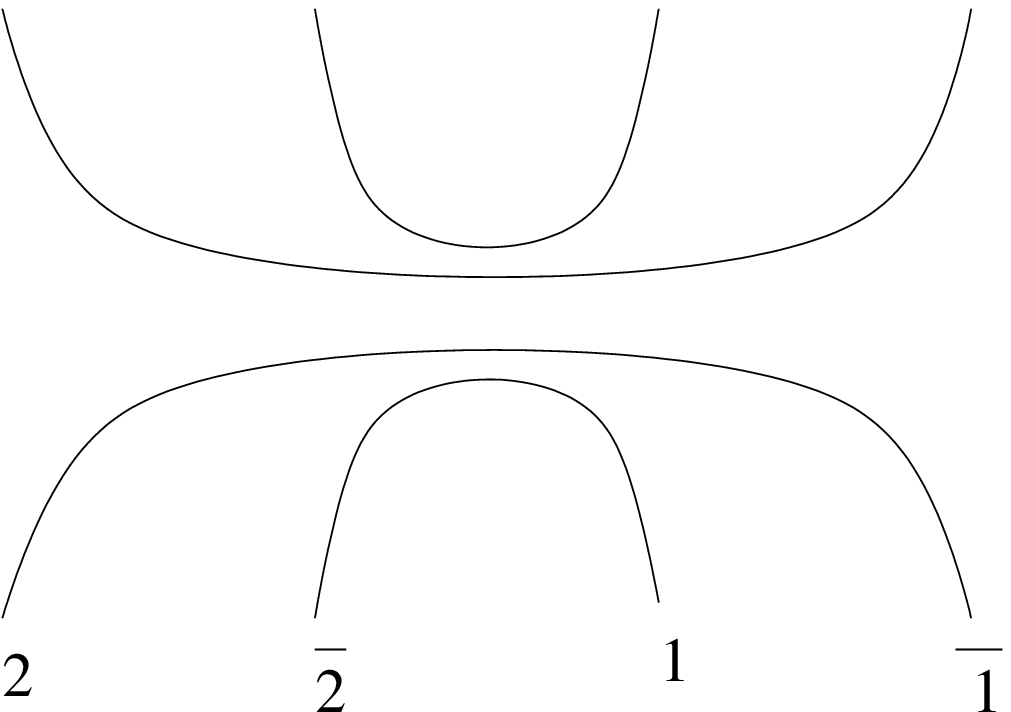}
\end{picture}}
\]
\caption{\label{fig14} $U (N_{f})_1 \times U (N_{f})_2
\times U (N_{f})_{\bar{1}} \times
U (N_{f})_{\bar{2}} \rightarrow U (N_{f})_{\rm{diag} (1,\bar{2})} \times
U (N_{f})_{\rm{diag} (2,\bar{1})}$.
}
\end{center}
\end{figure}

     From the structure of the inequalities (\ref{cri}) and
\C{ineq2}, we expect that at an arbitrary coupling the vacuum is in a phase
determined by an inequality of the form
\begin{equation}
f\Big( \frac{\lambda}{L_{1 \bar{1}}} \Big) + f
\Big(\frac{\lambda}{L_{2 \bar{2}}} \Big)  \ >  \ f \Big(
\frac{\lambda}{L_{1 \bar{2}}}\Big) + f\Big(
\frac{\lambda}{L_{2\bar{1}}} \Big),
\end{equation}
where $f(x)$ is a monotonically increasing function of $x$, and
\begin{displaymath}
f (x) \rightarrow \left\{ \begin{array}{ll}
e^{-2/x}, &{\rm as} ~~x \rightarrow 0, \\
x^4, & {\rm as} ~~x \rightarrow \infty.
\end{array} \right.
\end{displaymath}

\subsection{Four flavor branes spanning two dimensions}

So far we have considered stacks of flavor branes that lie along a straight line in $\mR^3$. In order to
understand the general patterns of chiral symmetry breaking, it is useful to look at flavor brane
configurations that are not collinear. With this in mind, we analyze chiral symmetry breaking in
the two simplest D-brane configurations spanning an $\mR^2$ in $\mR^3$, both at weak and strong coupling.  
These two configurations are given by two stacks of D6-branes and two stacks of
$\overline{\td6}$-branes lying in a plane in $\mR^3$, such that they form a rectangle. There are two
distinct orderings of the D-branes (upto charge conjugation), as we discuss below.

\subsubsection{$\td6-\td6-\overline{\td6}-\overline{\td6}$ rectangle}

We consider the D6 and $\overline{\td6}$-branes as depicted in figure \C{fig15} forming a rectangle
in $\mR^2$. In the diagram, each stack of D6 ($\overline{\td6}$)-branes is represented
by a point, given by a vertex of the rectangle.

\begin{figure}[ht]
\begin{center}
\[
\mbox{\begin{picture}(120,125)(0,0)
\includegraphics[scale=.5]{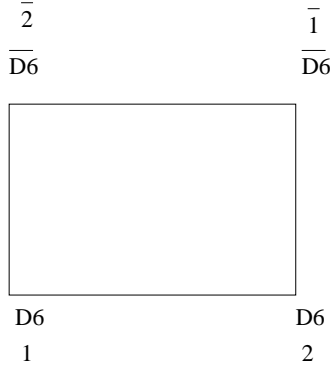}
\end{picture}}
\]
\caption{\label{fig15} $\td6-\td6-\overline{\td6}-\overline{\td6}$ rectangle.
}
\end{center}
\end{figure}

Proceeding as before, at weak coupling, the non--vanishing condensates are given by

\begin{equation}
 \vert \phi_{1\bar{2}} \vert
= \mu e^{ -L_{1 \bar{2}}/ \lambda} = \mu e^{ -L_{2 \bar{1}}/ \lambda} =\vert \phi_{2\bar{1}} \vert,
      \ \ \phi_{1\bar{1}}=0, \ \  \phi_{2\bar{2}}  = 0 .\
\end{equation}
Similarly in the strong coupling limit, there are two wormholes connecting the corresponding
vertices of the rectangle.

\subsubsection{ $\td6-\overline{\td6}-\td6-\overline{\td6}$ rectangle}

\begin{figure}[ht]
\begin{center}
\[
\mbox{\begin{picture}(120,125)(0,0)
\includegraphics[scale=.5]{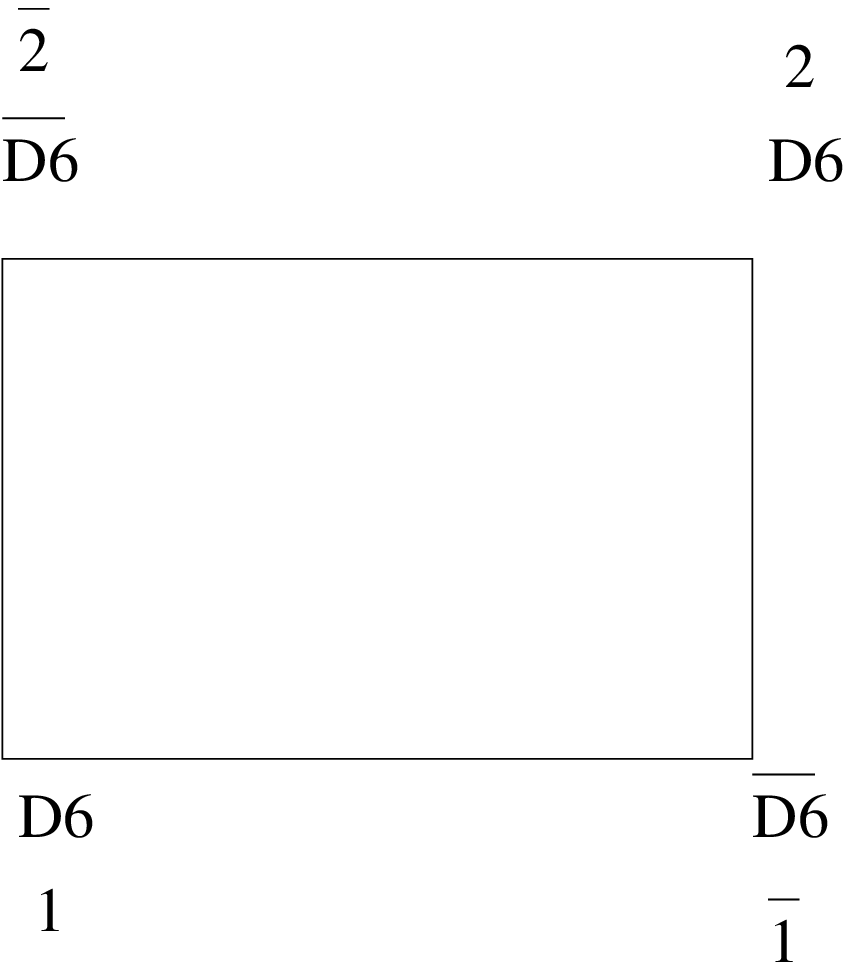}
\end{picture}}
\]
\caption{\label{fig16} $\td6-\overline{\td6}-\td6-\overline{\td6}$ rectangle.
}
\end{center}
\end{figure}

Finally, we consider the other possible configuration of D6 and $\overline{\td6}$-branes which form a
rectangle in $\mR^2$ as depicted in figure \C{fig16}. At weak coupling, if $L_{1\bar{1}} > L_{1\bar{2}}$,
the non--vanishing condensates are given by

\begin{equation}
 \vert \phi_{1\bar{2}} \vert
= \mu e^{ -L_{1 \bar{2}}/ \lambda} = \mu e^{ -L_{2 \bar{1}}/ \lambda} =\vert \phi_{2\bar{1}} \vert,
      \ \ \phi_{1\bar{1}}=0, \ \  \phi_{2\bar{2}}  = 0 .\
\end{equation}
On the other hand, if $L_{1\bar{2}} > L_{1\bar{1}}$, the non--vanishing condensates are given by

\begin{equation}
 \vert \phi_{1\bar{1}} \vert
= \mu e^{ -L_{1 \bar{1}}/ \lambda} = \mu e^{ -L_{2 \bar{2}}/ \lambda} =\vert \phi_{2\bar{2}} \vert,
      \ \ \phi_{1\bar{2}}=0, \ \  \phi_{2\bar{1}}  = 0 .\
\end{equation}
At strong coupling, there are corresponding wormhole solutions along the corresponding edges of the
rectangle in an obvious way.

Now when $L_{1\bar{1}} = L_{1\bar{2}}$, the D-brane configuration has an enhanced $U (2 N_f)_L \times
U (2 N_f)_R$ chiral symmetry classically, which gets dynamically broken to
$U (N_f)_{\rm{diag} (L,R)} \times U (N_f)_{\rm{diag} (L,R)}$ both at weak and strong coupling.

\section{General patterns of chiral symmetry breaking and restoration at high temperature}

 From the above examples, one can see the general patterns of chiral symmetry breaking both at
weak coupling as well as at strong coupling. In the weak coupling limit, in the single gluon
exchange approximation, we always get a generalized non--local GN model. Furthermore restricting to
length scales much larger that the separations between the flavor branes, we end up
getting local GN models. Interactions in these
models depend only on the distances $L_{i\bar{j}}$ between a stack of D6-branes
and another stack of
$\overline{\td6}$-branes placed at $L_i$ and $L_{\bar{j}}$ respectively,
through the corresponding GN coupling
$\lambda/L_{i\bar{j}}$. In particular, the coupling between a D6-$\overline{\td6}$ pair does not
depend on the angular orientation of the two stacks in $\mR^3$.
Thus in general, we will have generalized GN models with all possible couplings, and
the vacuum configuration will be determined by
the energetics; namely, the vacuum configuration will have non--vanishing condensates
of only those fermion bilinears connecting the D6--$\overline{\td6}$ pairs such that the energy 
is at its global minimum. 
On the basis of the constructions we have described, we expect that in any D-brane configuration 
with equal number of stacks of D6 and $\overline{\td6}$-branes, the vacuum configuration will
always have non--vanishing condensates such that all the fermions get mass dynamically. 
We also expect that if there are $P$ stacks of D6-branes
and $Q$ stacks of $\overline{\td6}$-branes such that $P \neq Q$, the vacuum configuration
will have $ \vert P- Q \vert$ stacks of unpaired D6 ($\overline{\td6})$ branes, depending on
whether $P > Q (P < Q)$. For certain D-brane configurations, there can be enhanced chiral symmetries.
For example, if a stack of D6-branes is equidistant from $K$ stacks of
$\overline{\td6}$-branes in $\mR^3$, then the configuration has an enhanced $U(N_f)_L \times U(K N_f)_R$
chiral symmetry, which is dynamically broken to $U (N_f)_{{\rm{diag}} (L,R)} \times U((K-1) N_f)_R$.
At weak coupling, one has a condensate given by a fermion bilinear involving the left--moving fermions 
from the D6-brane stack and the right--moving fermions from any one of the $K$ $\overline{\td6}$-brane
stacks. In the strong coupling limit, the analogues of condensates are wormholes. The vacuum 
configuration is determined by the the energetics of wormholes connecting 
pairs of D6-branes and $\overline{\td6}$-branes.

We have so far analyzed the chiral symmetry breaking which occurs at infinite $N_c$ at zero
temperature. If we now start heating the system to higher and higher temperatures, the chiral
symmetry gets restored as we now explain. In the GN model, it is well known that the chiral symmetry is
restored at a temperature $T_c$ given by~\cite{Jacobs:1974ys,Harrington:1974tf,Dashen:1974xz}
\be T_c = 0.57 m_f,\ee
where $m_f$ is the dynamically generated mass for the fermions at zero temperature. In fact,
this phase transition is second order in nature. Thus in the weak coupling limit where we obtain the
generalized GN models, there is chiral symmetry restoration at temperatures given by
\be \label{tempweak}
T_c (L_{i\bar{j}}) = 0.57 \mu e^{- L_{i\bar{j}} /\lambda (\mu)}.\ee
So as we heat up the system from zero temperature, the condensates start evaporating as soon as the
corresponding phase transition temperatures are attained. Thus at a sufficiently high temperature, all the
symmetries are restored. The strong coupling analysis is similar, with the phase transition temperature
given by~\cite{Antonyan:2006qy}~\footnote{This has been analyzed for the $D4-D8-\overline{D8}$ case
in~\cite{Aharony:2006da,Parnachev:2006dn}.}
\be T_c (L_{i\bar{j}}) \approx \frac{0.205}{L_{i\bar{j}}}.\ee
However, now the phase transition is first order in nature.

Thus we see that the D-brane configurations we have discussed above exhibit interesting patterns of chiral
symmetry breaking. The construction of generalized GN models from string theory opens up
several directions which might be worth looking at. It would be nice to understand the spectrum 
of bound states, solitons and their interactions in these theories both at weak and strong 
coupling, and possibly also at finite values of the coupling. Also one might be interested in analyzing
the integrability of these theories to compute the S--matrix. It might also be useful
to analyze configurations where there are directions transverse to both the color and the
flavor branes, which enable us to give tunable masses to the fermions. These
techniques are generalizable to other D-brane configurations and might be useful in
understanding generalized models of chiral symmetry breaking in other dimensions,
thus generalizing the results in~\cite{Antonyan:2006pg}. Finally, one can also analyze
generalized models of ${\rm{QCD}}_2$ by wrapping the D4-branes on 
$T^3$~\cite{Antonyan:2006qy}\footnote{A construction of ${\rm{QCD}}_2$ using a different
D-brane configuration has been done in~\cite{Gao:2006up}.} as mentioned before.

\section*{Acknowledgements}
We would like to thank S.~Giddings, J.~Maldacena, P.~Ouyang, and
J.~Polchinski for useful discussions. The work of A.~B. and A.~M.  
is supported by NSF Grant No.~PHY-0503584, and DOE Grant 
No.~DE-FG02-91ER40618 respectively.


\begin{thebibliography}{10}

\bibitem{Nambu:1961tp}
Y.~Nambu and G.~Jona-Lasinio, ``Dynamical model of elementary particles based
  on an analogy with superconductivity. I,'' {\em Phys. Rev.} {\bf 122} (1961)
345--358.

\bibitem{Gross:1974jv}
D.~J. Gross and A.~Neveu, ``Dynamical symmetry breaking in asymptotically free
  field theories,'' {\em Phys. Rev.} {\bf D10} (1974)
3235.

\bibitem{Coleman:1973jx}
S.~R. Coleman and E.~Weinberg, ``Radiative corrections as the origin of
  spontaneous symmetry breaking,'' {\em Phys. Rev.} {\bf D7} (1973)
1888--1910.

\bibitem{Witten:1978qu}
E.~Witten, ``Chiral symmetry, the 1/N expansion, and the SU(N) Thirring
  model,'' {\em Nucl. Phys.} {\bf B145} (1978)
110.

\bibitem{Affleck:1985wa}
I.~Affleck, ``On the realization of chiral symmetry in (1+1)-dimensions,'' {\em
  Nucl. Phys.} {\bf B265} (1986)
448.

\bibitem{Witten:1998zw}
E.~Witten, ``Anti-de Sitter space, thermal phase transition, and confinement in
  gauge theories,'' {\em Adv. Theor. Math. Phys.} {\bf 2} (1998) 505--532,
\href{http://www.arXiv.org/abs/hep-th/9803131}{{\tt hep-th/9803131}}.

\bibitem{Sakai:2004cn}
T.~Sakai and S.~Sugimoto, ``Low energy hadron physics in holographic QCD,''
  {\em Prog. Theor. Phys.} {\bf 113} (2005) 843--882,
\href{http://www.arXiv.org/abs/hep-th/0412141}{{\tt hep-th/0412141}}.

\bibitem{Sakai:2005yt}
T.~Sakai and S.~Sugimoto, ``More on a holographic dual of QCD,'' {\em Prog.
  Theor. Phys.} {\bf 114} (2006) 1083--1118,
\href{http://www.arXiv.org/abs/hep-th/0507073}{{\tt hep-th/0507073}}.

\bibitem{Karch:2002sh}
A.~Karch and E.~Katz, ``Adding flavor to AdS/CFT,'' {\em JHEP} {\bf 06} (2002)
  043,
\href{http://www.arXiv.org/abs/hep-th/0205236}{{\tt hep-th/0205236}}.

\bibitem{Son:2003et}
D.~T. Son and M.~A. Stephanov, ``QCD and dimensional deconstruction,'' {\em
  Phys. Rev.} {\bf D69} (2004) 065020,
\href{http://www.arXiv.org/abs/hep-ph/0304182}{{\tt hep-ph/0304182}}.

\bibitem{Kruczenski:2003uq}
M.~Kruczenski, D.~Mateos, R.~C. Myers, and D.~J. Winters, ``Towards a
  holographic dual of large-N(c) QCD,'' {\em JHEP} {\bf 05} (2004) 041,
\href{http://www.arXiv.org/abs/hep-th/0311270}{{\tt hep-th/0311270}}.

\bibitem{Babington:2003vm}
J.~Babington, J.~Erdmenger, N.~J. Evans, Z.~Guralnik, and I.~Kirsch, ``Chiral
  symmetry breaking and pions in non-supersymmetric gauge / gravity duals,''
  {\em Phys. Rev.} {\bf D69} (2004) 066007,
\href{http://www.arXiv.org/abs/hep-th/0306018}{{\tt hep-th/0306018}}.

\bibitem{Antonyan:2006vw}
E.~Antonyan, J.~A. Harvey, S.~Jensen, and D.~Kutasov, ``NJL and QCD from string
  theory,''
\href{http://www.arXiv.org/abs/hep-th/0604017}{{\tt hep-th/0604017}}.

\bibitem{Bak:2004nt}
D.~Bak and H.-U. Yee, ``Separation of spontaneous chiral symmetry breaking and
  confinement via AdS/CFT correspondence,'' {\em Phys. Rev.} {\bf D71} (2005)
  046003,
\href{http://www.arXiv.org/abs/hep-th/0412170}{{\tt hep-th/0412170}}.

\bibitem{Volkov:2005kw}
M.~K. Volkov and A.~E. Radzhabov, ``Forty-fifth anniversary of the
  Nambu-Jona-Lasinio model,''
\href{http://www.arXiv.org/abs/hep-ph/0508263}{{\tt hep-ph/0508263}}.

\bibitem{Antonyan:2006qy}
E.~Antonyan, J.~A. Harvey, and D.~Kutasov, ``The Gross-Neveu model from string
  theory,''
\href{http://www.arXiv.org/abs/hep-th/0608149}{{\tt hep-th/0608149}}.

\bibitem{Klimenko:1985ss}
K.~G. Klimenko, ``Generalization of Gross-Neveu model to the case of several
  coupling constants,'' {\em Theor. Math. Phys.} {\bf 66} (1986)
252.

\bibitem{Itzhaki:1998dd}
N.~Itzhaki, J.~M. Maldacena, J.~Sonnenschein, and S.~Yankielowicz,
  ``Supergravity and the large N limit of theories with sixteen supercharges,''
  {\em Phys. Rev.} {\bf D58} (1998) 046004,
\href{http://www.arXiv.org/abs/hep-th/9802042}{{\tt hep-th/9802042}}.

\bibitem{Jacobs:1974ys}
L.~Jacobs, ``Critical behavior in a class of O(N) invariant field theories in
  two-dimensions,'' {\em Phys. Rev.} {\bf D10} (1974)
3956.

\bibitem{Harrington:1974tf}
B.~J. Harrington and A.~Yildiz, ``Restoration of dynamically broken symmetries
  at finite temperature,'' {\em Phys. Rev.} {\bf D11} (1975)
779.

\bibitem{Dashen:1974xz}
R.~F. Dashen, S.-k. Ma, and R.~Rajaraman, ``Finite temperature behavior of a
  relavistic field theory with dynamical symmetry breaking,'' {\em Phys. Rev.}
  {\bf D11} (1975)
1499.

\bibitem{Aharony:2006da}
O.~Aharony, J.~Sonnenschein, and S.~Yankielowicz, ``A holographic model of
  deconfinement and chiral symmetry restoration,''
\href{http://www.arXiv.org/abs/hep-th/0604161}{{\tt hep-th/0604161}}.

\bibitem{Parnachev:2006dn}
A.~Parnachev and D.~A. Sahakyan, ``Chiral phase transition from string
  theory,''
\href{http://www.arXiv.org/abs/hep-th/0604173}{{\tt hep-th/0604173}}.

\bibitem{Antonyan:2006pg}
E.~Antonyan, J.~A. Harvey, and D.~Kutasov, ``Chiral symmetry breaking from
  intersecting D-branes,''
\href{http://www.arXiv.org/abs/hep-th/0608177}{{\tt hep-th/0608177}}.

\bibitem{Gao:2006up}
Y.-h. Gao, W.-s. Xu, and D.-f. Zeng, ``NGN, QCD(2) and chiral phase transition
  from string theory,'' {\em JHEP} {\bf 08} (2006) 018,
\href{http://www.arXiv.org/abs/hep-th/0605138}{{\tt hep-th/0605138}}.

\end{thebibliography}




\providecommand{\href}[2]{#2}\begingroup\raggedright\endgroup

\end{document}